\documentclass[aps, floatfix, prb, a4paper, twocolumn]{revtex4-2}
\usepackage{amsmath}
\usepackage{graphicx}
\usepackage{bm}
\usepackage{xcolor}
\usepackage[caption=false]{subfig}

\newcommand{\bpm}{\begin{pmatrix}}
\newcommand{\epm}{\end{pmatrix}}

\newcommand{\e}{\mathrm{exp}\!}

\begin{document}

\title{Half Quantum Vortices in Nematic Superconductor}

\date{\today}
\author{Pye Ton How}
\affiliation{Physics Division, National Center for Theoretical Sciences, Hsinchu 300, Taiwan}

\author{Sung-Kit Yip}
\affiliation{Institute of Physics, Academia Sinica, Taipei 115,
	Taiwan}
\affiliation{Institute of Atomic and Molecular Sciences, Academia
	Sinica, Taipei 106, Taiwan}
\affiliation{Physics Division, National Center for Theoretical Sciences, Hsinchu 300, Taiwan}

\begin{abstract}
Motivated by the superconductivity of $M_x$Bi$_2$Se$_3$, we study topological excitations in a nematic superconductor using Ginzburg-Landau theory.  An isolated excitation at low field is shown to be either a distorted phase vortex or a tightly-bounded pair of half quantum vortices.  Close to upper critical field $H_{c2}$, the vortex lattice is shown to be always hexagonal in the extreme type-II limit.  Due to the different symmetries of the vortex lattice states, at least two phase transitions must take place when the external field is lowered from $H_{c2}$.
\end{abstract}

\maketitle

\section{Introduction}

Superconductivity of $M_x$Bi$_2$Se$_3$ has attracted much attention\cite{Yonezawa2018}.  Observations of in-plane anisotropy\cite{Matano2016, Pan2016, Asaba2017, Yonezawa2017, Smylie2018, Willa2018, Sun2019} lead to suggestion that the superconducting order parameter is two-component and spontaneously breaks lattice rotational symmetry\cite{Fu2014, Venderbos2016}.  However, the leading theoretical candidate (odd-parity $E_u$ representation) is necessarily topological\cite{Fu2014}, yet experimental search for Majorana bound states turned out mixed results \cite{Sasaki2011, Levy2013, Peng2013}.  Besides, the current theoretical model calls for a pre-exisiting pinning field, and simple anisotropy measurements cannot rule out an anisotropic s-wave model as the alternative\cite{Venderbos2016, How2019}.  By theoretically investigating the topological excitations of a nematic superconductor, we seek to propose yet another way to prove (or reject) this ``nematic hypothesis''.

Half quantum vortex (HQV) is predicted to exist in a superfluid with a multi-component order parameter\cite{Volovik1976, Salomaa1985, Salomaa1987}.  It was also discussed in the context of a chiral superconductor without spin-orbit coupling\cite{Kee2000, DasSarma2006, Chung2007}.  In contrast with the conventional phase vortex (PV), where the phase of the order parameter winds by $2\pi$ around the core, the phase and \emph{orientation} of the order parameter each winds by $\pi$ around an HQV, resulting in a single-valued wavefunction topologically distinct from a PV.  In a superconductor (charged superfluid), an HQV carries only half a magnetic flux quantum.  Evidence of HQV has been reported in the polar phase of $^3$He\citep{Autti2016}, ultracold BEC\cite{Lagoudakis2009, Manni2012, Liu2015, Seo2015}, and superconducting Sr$_2$RuO$_4$\cite{Jang2011}.  Numerical solution of Ginzburg-Landau equation shows that tightly-bounded HQV pairs may exist in $M_x$Bi$_2$Se$_3$ at low applied field \cite{Zyuzin2017}.  Due to gauge screening and spin-orbit coupling unique to unconventional superconductors, the HQV physics is expected to be quite different from the counterpart in neutral superfluid.

In this work, we study the Ginzburg-Landau (GL) theory of an unconventional superconductor with a two-component order parameter, and analyze its topological excitations at both low field and near $H_{c2}$.  The theoretical model is introduced in sec II.  In section III we examine the stability of an isolated HQV pair at low field, and show that the pair can be the preferred topological excitation for certain range of parameters.  In section IV we explore the physics near $H_{c2}$.  We present a complete extension of the Abrikosov vortex lattice\cite{Abrikosov1957, Zhitomirskii1989, DeGennes1999} to the multi-component case, which to our knowledge has never been completely carried out in existing literature: previous authors\cite{Garg1994, Agterberg1998} employed various additional approximations to skirt around the issue of evaluating the microscopic magnetic field.  We find a ``vortex lattice'' solution where each unit cell carries the same topological charges identical of a low-field HQV pair.  In sec V, we explore how the low- and high-field limits can be connected based on their respective symmetry.  We argue that at least two phase transitions are required at intermediate fields to connect the two limits.  This is an unavoidable consequence if the superconductor has a multi-component, nematic order parameter, and we hope that our result helps to point a new direction for future experimental work toward the unequivocal identification of the superconductivity in $M_x$Bi$_2$Se$_3$.

\section{Ginzburg-landau Theory}

We assume the system is uniform along the $c$-direction of the lattice, thus effectively \emph{two-dimensional}.  The nematic superconductivity is modeled by the two-component order parameter $\eta_{i}$ (indices $i, j = x, y$) that transformed as a two-dimensional representation under the lattice point group ($D_{3d}$ for $M_x$Bi$_2$Se$_3$), i.e. to the naive eyes $\vec{\eta} = (\eta_x, \eta_y)$ transforms like a vector under spatial rotation.  However, one may identify $\vec{\eta}$ with $-\vec{\eta}$, since they effectively differ only by a global phase shift of $\pi$.  It is therefore best to think of $\vec{\eta}$ as a nematic director.  To quartic order in $\eta_i$, the general GL free energy density that respects $D_{3d}$ turns out to be isotropic.  The theory is minimally coupled to the 2D gauge field $A_i$.  These considerations result in the free energy density
\begin{widetext}
\begin{equation}
\mathcal{F} =
K_1 (p_i \eta_j)^{*}(p_i \eta_j)
+ K_2(p_i \eta_i)^{*}(p_j \eta_j)
+ K_3(p_i \eta_j)^{*}(p_j \eta_i)
+ \alpha (\eta_{i}^{*}\eta_{i})
+ \frac{\beta_1}{2} \left [(\eta_{i}^{*}\eta_{i})^2
+ \beta \vert\eta_{i} \eta_{i}\vert^2 \right]
+ \frac{(\nabla\!\times\!\vec{A})^{2}}{8\pi},
\label{GLFreeEnergy}
\end{equation}
where $p_i = -i \partial_i + A_i$.  We choose the unit $\hbar = (-e^{*}/c )= 1$.

It is often useful to consider the alternative basis $\eta_{\pm} = \eta_x \pm i \eta_y$.  Let $p_{\pm} = p_x \pm i p_y$ likewise, and one has
\begin{equation}
\begin{split}
\mathcal{F} &=
	\frac{K_{13}}{4}\left( \vert p_+ \eta_+ \vert^2 + \vert p_- \eta_- \vert^2 \right)
	+ \frac{K_{12}}{4}\left( \vert p_+ \eta_- \vert^2 + \vert p_- \eta_+ \vert^2 \right)
	+ \frac{\alpha}{2}\left(\vert\eta_+\vert^2
	+ \vert\eta_-\vert^2 \right)
	+ \frac{\beta_1}{8}\left(\vert\eta_+\vert^4
	+ \vert\eta_-\vert^4 \right)
	+ \frac{(\nabla\!\times\!\vec{A})^{2}}{8\pi} \\
& \quad \, + \frac{K_{23}}{4} \left[
	(p_+ \eta_-)^{*}(p_- \eta_+)
	+ (p_- \eta_+)^{*}(p_+ \eta_-) \right]
+ \frac{\beta_1}{2} \, \delta\beta \, \vert \eta_+\vert^2 \vert \eta_-\vert^2 ,
\end{split}
\label{freeEnergy+-}
\end{equation}
\end{widetext}
where $K_{ij}$ stands for $K_i+K_j$, and $\delta \beta = 1/2 + \beta$.

A nematic director specify a special orientation in space.  In two dimensions, however, the orientation orthogonal to it is also uniquely defined.  One can therefore expect the existence of a dual theory where $\vec{\eta}$ is rotated by $\pi/2$.  Indeed, under the following duality transformation, the free energy stays invariant:
\begin{equation}
\begin{split}
\eta_i \rightarrow \epsilon_{ij} \eta_j, &\qquad
K_1 \rightarrow K_1 + K_2 + K_3, \\
K_2 \rightarrow -K_3, &\qquad
K_3 \rightarrow -K_2.
\end{split}
\label{dualityFull}
\end{equation}
where $\epsilon_{ij}$ is the Levi-Civita Symbol.

In the absence of external gauge field, this theory has two distinct phases.  These uniform solutions are of the form:
\begin{equation}
\vec{\eta} = (\eta_x, \eta_y) \propto
	\begin{cases}
		e^{i \chi}(1, \pm i),  &\beta > 0, \text{chiral} \\
		e^{i \chi}(\cos \theta, \sin \theta),  &0 > \beta > -1,\text{nematic}
	\end{cases}
\label{uniformState}
\end{equation}
The chiral phase preserves rotational symmetry, but breaks time-reversal, while the nematic phase is time-reversal invariant, but spontaneously breaks the rotational symmetry of the lattice.  We will focus on the nematic phase, as it is the theoretical candidate for superconductivity in $M_x$Bi$_2$Se$_3$.

We will set $K_2 = K_3$ for the reminder of this paper.  Take the model \eqref{GLFreeEnergy} and integrate by part, it can be seen that the difference $(K_2 - K_3)$ only couples to $(\nabla \!\times\!\vec{A}) (\vec{\eta})^{*} \!\times\!\vec{\eta}$.  The cross product $(\vec{\eta})^{*} \!\times\!\vec{\eta}$ is odd under time reversal, and traces its origin in Fermi liquid theory to particle-hole asymmetry\cite{Ambegaokar1973, Salomaa1987}.  We therefore expect it to be small\footnote{One expects particle-hole near-symmetry around a Fermi surface: coefficients to asymmetric terms are of the order $(T_c/E_F)^2 \approx 10^{-6}$\cite{Leggett1975} \cite{*[{Also see the supplemental material of }][{.}] Venderbos2016a}, using $E_F \approx 0.25eV$\cite{Wray2010} and $T_c \approx 3K$\cite{Hor2010, Kriener2011}}.  We also argue that relaxing the $K_2 = K_3$ condition does not change our result in any qualitative way\footnote{Near $H_{c2}$, the phase diagram in \cite{Zhitomirskii1989} implies that the model remains in the same phase when $K_2-K_3$ slightly deviates from zero, and we expect no drastic change in its behavior.  As the term is directly proportional to $\nabla \times \vec{A}$, we expect it to be even less important at low field.}.

Zhitomirskii\cite{Zhitomirskii1989} investigated the behavior of this model.  It will be helpful to define $C = (K_2+K_3)/2K_1$, following his notation \footnote{The value of $C$ can be estimated from the anisotropic ratio of in-plane $H_{c2}$; see \cite{Hess1989} and supplemental material of \cite{Willa2018}.  For Cu doping, we find $C \approx 0.6$ using \cite{Yonezawa2017}.  For Sr however, the reported ratios in \cite{Pan2016, Smylie2018, Willa2018, Sun2019} all leads to $C > 1$, violating the stability bound.}.  Positivity of the free energy leads to the stability conditions
\begin{equation}
\beta_1 > 0,\quad K_1 > 0,\quad 1 > C >-1/3,\quad \beta > -1.
\end{equation}

Experimentally reported ratio of coherence length to penetration depth $\xi/\lambda$ is typically of the order $O(100)$ \cite{Kriener2011, Shruti2015}, putting the material firmly in the extreme type-II limit.  Please see the appendix for relevant discussion.

\section{HQV pair at low field}

\subsection{Topology of HQV}

Consider the nematic solution in \eqref{uniformState}.  It has two angular parameters $\chi$ and $\theta$.  Topologically, a PV is a defect where only the phase $\chi$ winds around the vortex center, but the addition of $\theta$ opens up more exotic possibilities.  When viewed from spatial infinity, an HQV centered at the origin resembles:
\begin{equation}
\chi = n_p \frac{\phi}{2} + \delta\chi, \qquad
\theta = \theta_{\infty} + n_o \frac{\phi}{2} + \delta \theta,
\label{HQVform}
\end{equation}
where $\phi$ is the azimuthal angle, and $(n_p, n_o) = (\pm1, \pm1)$ marks its topological charge, associated respectively with the phase and orientation winding.  Here $\delta\chi$ and $\delta\theta$ are topologically-trivial deformations.  While the overall wavefunction remains single-valued, $\chi$ and $\theta$ each winds by only $\pi$ around the center.  Thus the name \emph{half} vortex.

In the $\pm$-basis introduced earlier, \eqref{HQVform} becomes
\begin{equation}
\begin{split}
(\eta_+, \eta_-) &\sim \left(e^{i\phi(n_p+n_o)/2},
\, e^{i\phi(n_p-n_o)/2}\right) \\
& = \left(e^{\pm i \phi}, 1 \right) \, \text{or} \,
	\left(1, e^{\pm i \phi}\right).
\end{split}
\label{HQV+-}
\end{equation}
A HQV can thus be viewed as a single vortex of one of the $\eta_{\pm}$ components, in a non-zero background of the other.  Note that in this language a PV has $(n_p = \pm 2,n_o = 0)$, and is topologically the same as two HQVs with opposite $n_o$ charges sitting at the same point:
\begin{equation*}
(\pm 1, 1) + (\pm 1, -1) \rightarrow (\pm 2, 0).
\end{equation*}

A superconductor exhibits the Meissner effect where the phase fluctuation is screened: the gradient energy due to the phase winding is localized around a topological excitation and remains finite.  In contrast, the energy cost of the orientation winding is divergent with the system size\cite{Babaev2002, Chung2007}.  In other words, an overall non-zero $n_o$ charge is energetically forbidden.  HQVs must therefore come in pairs with opposite $n_o$ charges, though the stability of the pair is yet unknown.  Does the pair reach equilibrium at finite separation, or does it collapse and merge into one PV?


\subsection{Extreme London Limit: Large Pair Separation}

To proceed, we consider in this part the extreme London limit, where the magnitude $\vert\vec{\eta}\vert$ is everywhere uniform except at the point-like core of a topological defect.  When the HQV pair is separated by a distance $d \gg \lambda$ the penetration depth, the gauge interaction is screened out, and the energy in the gauge sector is essentially independent of the separation.  On the other hand, the orientation gradient $\nabla \theta$ still significantly deviates from zero in the region between the two cores, and there is no screening effect for $\theta$.  Reducing $d$ evidently reduces the gradient energy.  The HQV pair is attractive at this very large separation.

We then consider the intermediate case, where the cores are still well separated, but the separation $d < \lambda$.  As $\vec{A}$ varies with length scale $\lambda$, it can be treated as approximately uniform.  We will set $\vec{A} = 0$ for this part, essentially considering a neutral superfluid.  In the London limit, free energy \eqref{GLFreeEnergy} reduces to:
\begin{equation}
\begin{split}
\mathcal{F} &= Z \Big \lbrace
\vert \nabla \theta \vert^2 + \vert \nabla \chi \vert^2
+ g \big [ \left(\hat{u} \times \nabla \theta \right)^2 +
\left(\hat{u} \cdot \nabla \chi \right)^2 \\
& \qquad  -\left(\hat{u} \cdot \nabla \theta \right)^2
- \left(\hat{u} \times\!\nabla\!\chi \right)^2
\big] \Big\rbrace,
\end{split}
\label{reducedFE}
\end{equation}
where $Z = K_1 \vert\alpha\vert(1+C)/[\beta_1(1+\beta)]$ gives overall dimensions, $g = C/(1+C)$, and $\hat{u} = (\cos \theta, \sin \theta)$ is the orientation unit vector.  The duality \eqref{dualityFull} becomes, for this effective theory,
\begin{equation}
\theta \rightarrow \theta + \frac{\pi}{2}, \qquad
Z \rightarrow Z, \qquad
g \rightarrow -g.
\end{equation}

The total free energy of the pair $F_{\text{pair}}$ can be split into three parts: the energy for two isolated HQV cores $2F_{\text{core}}$, the logarithmic interaction energy $F_{\text{log}}$, and an dipole potential $F_{\text{dip}}$ that depends on the relative alignment of the pair to the background $\vec{\eta}$ at infinity:
\begin{equation}
F_{\text{pair}} = 2F_{\text{core}} + F_{\text{dip}} + F_{\text{log}}.
\end{equation}
The interaction $\mathcal{F}_{\text{log}}$ is expected to be purely logarithmic in $d$ because \eqref{reducedFE} is scale-invariant, and it also determines the stability of the HQV pair.  Finding $\mathcal{F}_{\text{log}}$ will be our main goal here.  We will determine $\mathcal{F}_{\text{log}}$ by matching $F_{\text{pair}}$ with $F_{\text{PV}}$, the free energy of an isolated PV, in the limit of zero pair separation.

The effective free energy \eqref{reducedFE} has a very simple form in the limit of $g \rightarrow 0$.  We will calculate $F_{\text{core}}$, $F_{\text{dip}}$, and in addition the free energy of a single PV $F_{\text{PV}}$ perturbatively in $g$.

For a single, isolated HQV, we seek solution to the the Ginzburg-Landau equations based on \eqref{reducedFE} that has the desired topological winding \eqref{HQVform}.  The GL equations are solved for general values of $n_o$ and $n_p$.  To leading order in $g$ the solution turns out to be:
\begin{equation}
\begin{split}
\theta &= \theta_{\infty} + 
	\frac{n_o}{2} \phi + \frac{g}{4} \frac{n_o^2-4n_o+n_p^2}{(2-n_o)^2}
	\sin\left[(2-n_o)\phi - 2\theta_{\infty}\right] \\
\chi &=
	\frac{n_p}{2} \phi + \frac{g}{2} \frac{n_p}{(2-n_o)}
	\sin\left[(2-n_o)\phi -2\theta_{\infty}\right],
\end{split}
\label{HQVperturbative}
\end{equation}
where $\phi$ is the azimuthal angle and $(n_o, n_p) = (\pm 1, \pm 1)$ for an isolated HQV.   This solution is consistent with the duality, which rotates $\theta_{\text{HQV}}$ by $\pi/2$ on the left hand side, and sends $g \rightarrow -g$ and rotates $\theta_{\infty}$ by $\pi/2$ on the right hand side.  

The solution \eqref{HQVperturbative} is then inserted back into \eqref{reducedFE} to determine $F_{\text{core}}$:
\begin{equation}
F_{\text{core}} = Z \pi \log\left( \Lambda/\xi \right) \left( 1 - \frac{g^2}{2} + O(g^4) \right).
\label{HQVcore}
\end{equation}
The series contains only even powers of $g$ because the free energy is self-dual.  The spatial integral is formally divergent, and needs to be regularized with both short- and large-distance cutoffs $\xi$ and $\Lambda$.  The cutoff $\xi$ represents the small but finite core size.  On the other hand, $\Lambda$ reflects two distinct physical mechanisms: the gauge part of the gradient energy is screened and cut off beyond the magnetic penetration depth, while the orientation gradient energy is only cut off by the system size.  If another HQV is available to form a pair, this orientation gradient energy will be instead cut off by the pair separation.

\begin{figure}
\includegraphics[width=0.4\textwidth]{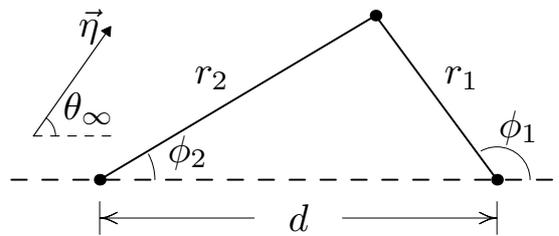}
\caption{\label{pairGeometry} The geometry of an HQV pair of separation $d$.  This defines the radii $r_1$, $r_2$, and the angles $\phi_1$, $\phi_2$.  The background $\vec{\eta}$ at spatial infinity is orientated at angle $\theta_{\infty}$ relative to the pair.}
\end{figure}

To calculate $F_{\text{dip}}$, consider a pair of HQVs centered at $(\pm d/2, 0)$ respectively.  See Fig \ref{pairGeometry}.  One only needs the zeroth-order solution to the GL equations here:
\begin{equation}
\begin{split}
\chi &= \frac{1}{2}(\phi_1 + \phi_2) + \dots \quad \\
\theta &= \theta_{\infty} + \frac{1}{2} (\phi_1 - \phi_2) + \dots
\end{split}
\label{pairForm}
\end{equation}
The dipole energy turns out to be
\begin{equation}
F_{\text{dip}} = Z \pi \cos(2\theta_{\infty}) \,  \left( -g \pi + O(g^3) \right).
\label{dipole}
\end{equation}
The self-dual requirement implies the absence of even powers of $g$.  This dipole potential aligns the pair to be parallel (perpendicular) to the background $\vec{\eta}$ when $C$ is positive (negative).

We will soon see that the interaction $F_{\text{log}}$ begins at $O(g^2)$, while $F_{\text{dip}}$ \eqref{dipole} starts at $O(g)$.  It is therefore sensible to assume that the pair always orients itself to minimize $F_{\text{dip}}$, which then becomes a mere additive constant.  Topologically, the pair is allowed to merge to form a single PV, and one therefore expects $F_{\text{pair}} \rightarrow F_{\text{PV}}$ in the limit of zero separation.  In particular, the coefficients of the logarithmic part on both sides must match.  This allows us to identify the interaction energy $F_{\text{log}}$ as the ``missing'' part when comparing $2F_{\text{core}}$ with $F_{\text{PV}}$.

To determine $F_{\text{PV}}$, one takes the solution \eqref{HQVform}, sets the topological charges to be $(n_p, n_o) = (\pm2, 0)$, and plugs the result into free energy \eqref{reducedFE} to get:
\begin{equation}
F_{\text{PV}} = Z \pi \log\left(\Lambda/\xi\right)
\left ( 2 - \frac{5}{4}g^2 + O(g^4) \right).
\label{Fpv}
\end{equation}
Comparison with $2F_{\text{core}}$ then leads to
\begin{equation}
F_{\text{log}} = Z  \pi \log\left( \Lambda/d \right )
	\left(-\frac{g^2}{4} + O(g^4)\right).
\label{interaction}
\end{equation}
The placement of $d$ in the logarithm is such that $F_{\text{pair}} \rightarrow F_{\text{PV}}$ when $d \approx \xi$.  This is an attraction: the pair continues to collapse at intermediate separation.

\subsection{When The Cores Overlap: Very Short Separation}

Finally, if the cores repel at very short distance, two HQVs can stay as a tightly-bounded pair; otherwise the pair is unstable and prone to collapse into a single PV.  We must now exit the London limit and allow for variation of order parameter magnitude within the vortex core.  An HQV here is interpreted as a single vortex of $\eta_+$ or $\eta_-$, and the geometry of the pair remains identical to Fig \ref{pairGeometry}.

To estimate the free energy cost, we employ the simple ansatz: two cylindrically-symmetric vortices of $\eta_+$ and $\eta_-$ placed side-by-side.  To be concrete, the order parameter configuration is taken to be:
\begin{equation}
\begin{split}
(\eta_+, \eta_-) &= \eta_{\infty} \left(e^{i(\phi_1 +\theta_{\infty})} f(r_1),
	e^{i(\phi_2 -\theta_{\infty})} f(r_2)\right), \\
f(r) &= \Theta\left(\frac{r}{\xi}-1\right) + \Theta\left(1-\frac{r}{\xi}\right) \, r/\xi,
\end{split}
\label{overlappingPair}
\end{equation}
where the amplitude $\eta_{\infty} = \sqrt{-\alpha/[2\beta_1(1+\beta)]}$.  The function $f(r)$ represents the amplitude variation around the core, starting out at zero at $r = 0$ and saturated when $r > \xi$.  We also assume $d \ll \xi$.

One must employ the full GL free energy \eqref{freeEnergy+-} directly, but only the terms that couple $\eta_+$ and $\eta_-$ will be dependent on the separation $d$.  Let us define two integrals:
\begin{equation}
\begin{split}
I_{23} &= \int \! d^2r \cos(2\phi_2)
\left[f'(r_2) - \frac{f(r_2)}{r_2}\right]
\left[f'(r_1) + \frac{f(r_1)}{r_1}\right], \\
I_{\beta} &= \int \! d^2r \, \left(f(r_1)^2 f(r_2)^2-1 \right).
\end{split}
\end{equation}
The $d$-dependent part of the free energy $F_{\text{int}}$ is given by
\begin{equation}
F_{\text{int}} = \frac{K_1 C}{2} \eta_{\infty}^2 \cos 2\theta_{\infty}\, I_{23}
+\beta_1\, \delta\beta \, \eta_{\infty}^4\, I_{\beta}.
\end{equation}

\begin{figure}
\includegraphics[width=0.4\textwidth]{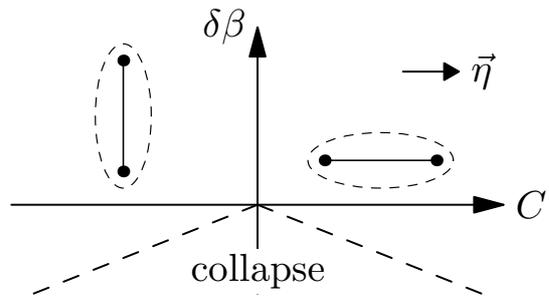}
\caption{\label{phaseDiagrams} The phase diagram of an isolated HQV pair at low field.  Depending on the sign of $C$, the pair is either oriented parallel or perpendicular to the background $\vec{\eta}$.  For $\delta\beta < 0$, the pair may collapse into one single PV if $\vert C \vert$ isn't big enough.}
\end{figure}

Both $I_{23}$ and $I_{\beta}$ are decreasing function of $d$.  To minimize $F_{\text{int}}$, the orientation $\theta_{\infty}$ will be such that $C \cos 2\theta_{\infty} = -\vert C \vert$, that is, the pair is aligned parallel (perpendicular) to the $\vec{\eta}$ at infinity if $C$ is positive (negative).  This is consistent with the dipole alignment found in section II B.  And then the $I_{23}$ term always favors separation.  The $I_{\beta}$ term, on the other hand, favors separation only if $\delta\beta > 0$.  If $\delta\beta < 0$, the competition between the two terms determines whether the pair collapses into one single PV.  The qualitative behavior is summarized in Fig \ref{phaseDiagrams}.

\section{Vortex Lattice at high field}

\subsection{Generalized Abrikosov Method}

After investigating the isolated vortex/pair solution, we switch gear to examine the solution at high field, where such topological defects exists at a finite density.  Near $H_{c2}$, the external magnetic field dominates, and there is no distinction between the nematic and chiral phases.  We return to the free energy \eqref{freeEnergy+-}, as it is natural to consider $\eta_{\pm}$ here, and seek a periodic vortex lattice solution that generalizes the seminal work of Abrikosov for s-wave type-II superconductor\cite{Abrikosov1957}.  Zhitomirskii\cite{Zhitomirskii1989} made an attempt in this direction, but incorrectly\cite{Lukyanchuk1993} concluded that the vortex lattice is always distorted from the hexagonal arrangement.

Our first step is to solve the linearized GL equation at magnetic field $\nabla \times \vec{A} = H = H_{c2}$:
\begin{widetext}
\begin{equation}
\begin{pmatrix}
(1+C/2) (p_+ p_- - H) + \alpha/K_1 & C \, (p_+)^2 \\
C \, (p_-)^2  &  (1+C/2) (p_+ p_- - H) + \alpha/K_1
\end{pmatrix}
\begin{pmatrix}
\eta_+ \\
\eta_-
\end{pmatrix}
= 0
\label{linearized}
\end{equation}

The value of $H_{c2}$ is determined by requiring that the equation has solutions at all.  The result, first due to Zhitomirskii\cite{Zhitomirskii1989}, adapted to our assumption $K_2 = K_3$, is
\begin{equation}
H_{c2} = \frac{\vert\alpha\vert}{K_1} \left[ 3(1+C) - 2\sqrt{1+2C+3C^2} \right].
\end{equation}
\end{widetext}

We look for a periodic solution with primitive lattice vectors $(a, 0)$ and $b (\cos \alpha, \sin \alpha)$.  These parameters satisfy the constraint $a b \sin\alpha = 2\pi/H_{c2}$, and it suffices to consider distinct $\rho = b/a$ and $\alpha$.  We adopt the Landau gauge $(A_x, A_y) = \left(-H_{c2} y, 0\right)$, define the Landau level wavefunctions:
\begin{equation}
\begin{split}
\phi_{0}^{(n)} &=
	\e \left[i \frac{2n\pi}{a} x 
	-\frac{H_{c2}}{2} \left( y - n b \sin \alpha \right)^{\!2} \right],\\
\psi_{j} &= \left(\frac{p_x + ip_y}{\sqrt{2H_{c2}}}\right)^j
	\sum_{n} \e\left[ i \pi \rho n(n-1) \cos \alpha \right] \phi_{0}^{(n)},
\label{LandauLevel}
\end{split}
\end{equation}
and finally write down the periodic ansatz that solves \eqref{linearized}:
\begin{equation}
\begin{split}
\eta_{+} &= \sqrt{\Omega}\, \psi_2, \qquad
\eta_{-} = \omega \, \sqrt{\Omega}\, \psi_0 \\
\omega &= \frac{2C}{-(1+C)+\sqrt{3C^2+2C+1}},
\end{split}
\label{ansatz}
\end{equation}
This will be employed as a variational ansatz, with $\rho$ and $\alpha$ as variational parameters.  The magnitude $\Omega$ is in principle another parameter, but in practice it is eliminated in the process.

We consider the Gibb's free energy density
\begin{equation}
\mathcal{G} = \mathcal{F}_2 + \mathcal{F}_4 + \left( \nabla \times \vec{A} - H \right)^2/8\pi.
\end{equation}
Here $\mathcal{F}_2$ and $\mathcal{F}_4$ are the parts of $\mathcal{F}$ that are quadratic and quartic in $\vec{\eta}$, respectively.  The applied $H$ field is allowed to be infinitesimally lower than $H_{c2}$: we define $\Delta H = (H_{c2} - H) > 0$.  But we still employ the same linear solution \eqref{ansatz} as variational ansatz, and set out solve for the gauge field configuration and minimize the total free energy.

To this end, the gauge field $\vec{A}$ is split into two parts:
\begin{equation}
\vec{A} = \vec{a} + \vec{A}_0
\end{equation}
where $\nabla \times \vec{A}_0 = H_{c2}$.  The field equation for $\vec{A}$ becomes:
\begin{equation}
\frac{\partial\mathcal{F}_2}{\partial \vec{A}} = -
	\frac{1}{4\pi}\nabla \times \left( \nabla\times\vec{a} + \Delta H\right).
\label{Maxwell}
\end{equation}
Assuming we can integrate this equation, we denote
\begin{equation}
h_s \equiv \nabla \times \vec{a} + \Delta H.
\label{hs}
\end{equation}
This assumption is not as innocuous as it seems.  For the s-wave case, the undoing of the curl is possible thanks to the property of the lowest Landau level wavefunction.  As \eqref{ansatz} contains higher Landau levels, the same trick would fail, and to our knowledge all previous authors employed various approximations to avoid the actual integration\cite{Garg1994, Agterberg1998}.  We point out that it is still possible to obtain an analytical expression for $h_s$, again thanks to the properties of the Landau levels.  We give the result here and leave the derivation to the appendix:
\begin{equation}
\begin{split}
\frac{h_s}{\pi K_1 \Omega} &=
	-(1+C) \vert\psi_2\vert^2
	- 2(2+2C + \omega C) \vert\psi_1\vert^2 \\
	& \qquad - [(4 + \omega^2)(1+C) + 6\omega C ] \vert \psi_0\vert^2.
\end{split}
\label{magneticField}
\end{equation}

The rest of the derivation stays parallel to the s-wave case.  Let the angular brackets $\langle \dots \rangle$ denotes spatial average.  Using the GL equation, one can deduce
\begin{equation}
\langle \mathcal{F}_2 \rangle + 2 \langle \mathcal{F}_4 \rangle = 0.
\end{equation}
However, since \eqref{ansatz} solves the linearized GL equation at $\vec{a} = 0$, we may schematically expand:
\begin{equation}
\langle \mathcal{F}_2 \rangle \approx \left\langle \vec{a} \cdot \frac{\partial \mathcal{F}_2}{\partial \vec{A}} \right\rangle
\end{equation}
The right hand side does not exactly makes sense, because $\vec{a}$ can't remain small for the entire space regardless of our gauge choice.  We employ \eqref{Maxwell} and integrate by part to cast it into a more sensible form:
\begin{equation}
-2 \langle \mathcal{F}_4 \rangle =
\langle \mathcal{F}_2 \rangle \approx
	- \frac{1}{4\pi} \left\langle (h_s - \Delta H) h_s\right\rangle.
\label{trick}
\end{equation}

With \eqref{magneticField} and \eqref{trick}, we finally have all necessary ingredients.  The Gibbs free energy can be cast into:
\begin{equation}
\begin{split}
\langle \mathcal{G} \rangle &= \langle \mathcal{F}_2 + \mathcal{F}_4 + \frac{1}{8\pi}(\nabla \times \vec{a} + \Delta H)^2 \rangle = -\frac{\Delta H^2}{8\pi} \frac{1}{R}, \\
R &= \frac{\langle \mathcal{F}_4 - h_s^2/8\pi \rangle}{\langle h_s \rangle^2/8\pi},
\end{split}
\label{Gibbs}
\end{equation}
The ratio $R$ is manifestly $\Omega$-independent.  Equation \eqref{Gibbs} is general, and yields exactly the Abrikosov formula when applied to an s-wave superconductor.

\subsection{Lattice Structure}

With analytic expressions of both $f_4$ and $h_s$, $R$ in \eqref{Gibbs} is numerically minimized with respect to $\rho$ and $\alpha$ to determine the most favorable lattice structure.

We introduce the ``GL parameter'' $\kappa$:
\begin{equation}
\kappa \equiv \frac{\beta_1}{8\pi K_1^2}  \, \frac{\min(1, \sqrt{1+\beta}) }{3(1+C)-2\sqrt{3C^2 + 2C + 1}},
\end{equation}
where $\kappa > 1/\sqrt{2}$ is the criterion for type-II behavior (see Appendix A for a detailed discussion).  The trio $(\kappa, C, \beta)$ are the dimensionless parameters that control the behavior of the system.

\begin{figure}
\includegraphics[width = 0.4\textwidth]{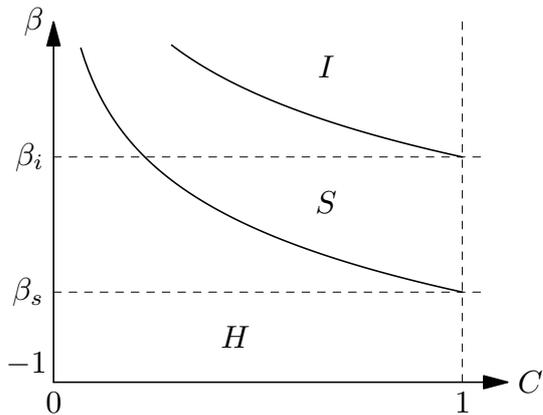}
\caption{\label{vortexLatticePhaseDiagram} The phase diagram for vortex lattice near $H_{c2}$ at constant $\kappa$.  Hexagonal, square and irregular phases are labelled as $H$, $S$ and $I$, respectively.  $\beta_s$ and $\beta_i$ are in general increasing function of $\kappa$.}
\end{figure}

Given the duality \eqref{dualityFull}, only $C \geq 0$ needs to be considered.  At $C = 0$, ansatz \eqref{ansatz} is effectively single-component, and hexagonal lattice is always favored, independent of $\beta$ and $\kappa$.  For $C > 0$, the phase diagram we obtained numerically is shown in Fig \ref{vortexLatticePhaseDiagram}.  In addition to hexagonal and square phases, we observed the so-called "irregular" phase, where the free energy landscape is extremely flat, with multiple competing near-degenerate minima, all of which have only oblique symmetry.  Both transitions appear sharp, and $\kappa$ does not affect the qualitative behavior.  $M_x$Bi$_2$Se$_3$ is estimated to have a $\kappa$ of $O(100)$\cite{Kriener2011, Shruti2015}, and has $\beta < 0$ for being in the nematic phase.  We found $\beta_s \approx 0.32$ and $\beta_i \approx 0.36$ for $\kappa = 100$, putting $M_x$Bi$_2$Se$_3$ in the hexagonal phase.

%
%

\begin{figure}
\includegraphics[width=0.4\textwidth]{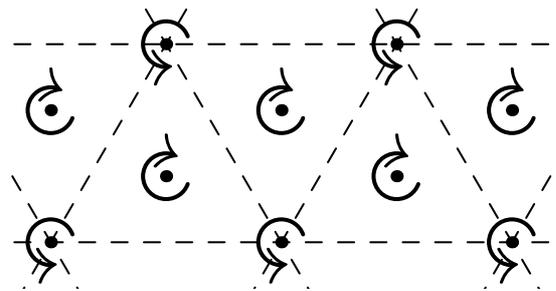}
\caption{\label{triangularLattice} The zeroes of the $\eta_+$ component in the hexagonal vortex lattice solution.  The arrows indicate the sense of phase winding, where counterclockwise is positive.  The intersection of dashed lines are the location of zeroes for the $\eta_-$ component; all of them have negative winding.}
\end{figure}

Assuming a hexagonal lattice, let us examine the structure of \eqref{ansatz}.  The $\eta_-$ component is identical to the hexagonal solution of an s-wave superconductor, and contains exactly one zero per unit cell, each with winding number $-1$ (clockwise).  In contrast, the $\eta_+$ component has \emph{three} zeroes per unit cell.  One of them coincides with the zero of $\eta_-$, and carries winding number $+1$ (counterclockwise).  The other two zeroes are located at the centers of the two equilateral triangles, respectively, and each has winding number $-1$.  The total winding number per unit cell for the $\eta_+$ component still sums to $-1$.  The relative phase between the $\eta_\pm$ components is periodic.  Considering both components, the total topological charge per unit cell is equal to that of one PV, or one HQV pair at low field.  See Fig \eqref{triangularLattice}.

\section{Bridging the two limits}

As the external field $H$ is lowered, the high-field vortex lattice must evolve into an array of well-isolated topological excitations.  These can be single PVs or tightly-bounded HQV pairs, as previously discussed.

\begin{figure}
\subfloat[][]{\includegraphics[width=0.17\textwidth]{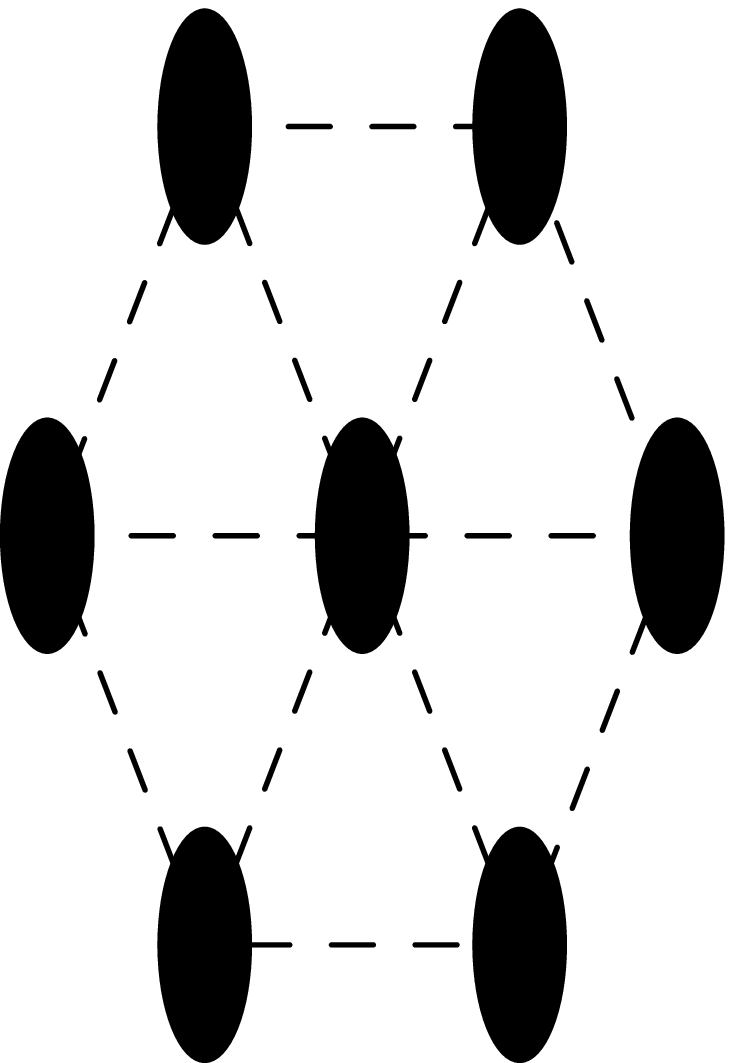}}
\qquad
\subfloat[][]{\includegraphics[width=0.17\textwidth]{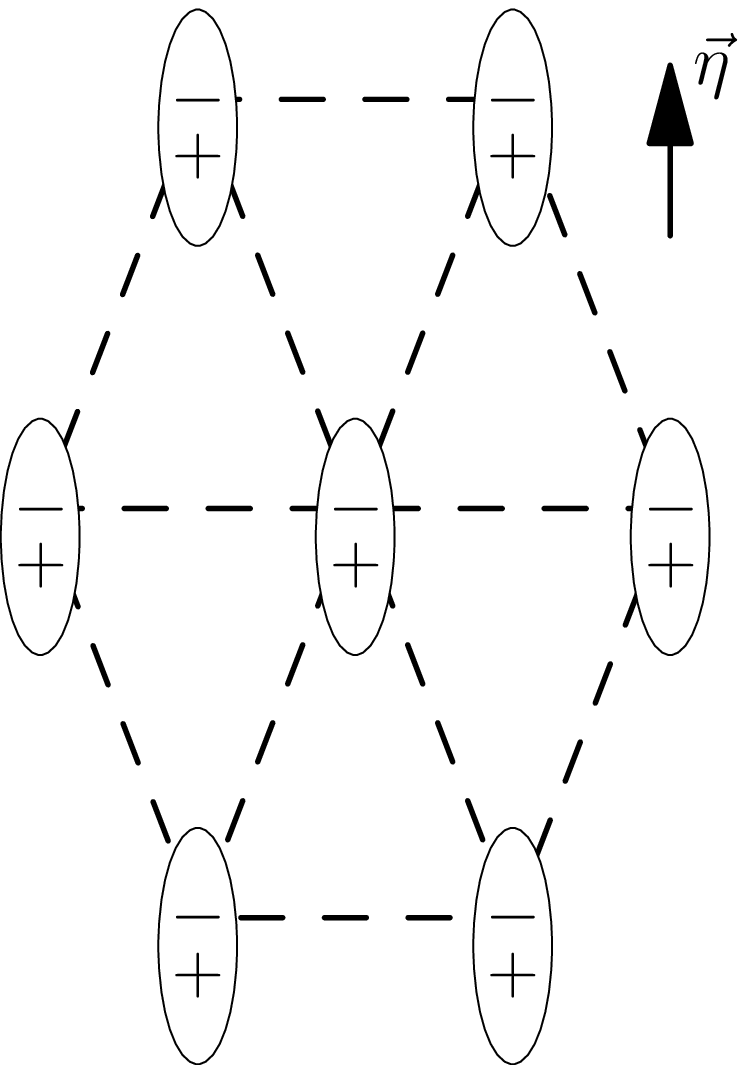}}
\caption{\label{lowField} Illustrations of the vortex lattice structure near $H_{c1}$.  (a) the $PV$ phase.  (b) the $HQV$ phase.  The elliptical vortices, the $HQV$ pair dipoles, and the distorted hexagonal lattice are all aligned with the background $\vec{\eta}$.}
\end{figure}

Near $H_{c2}$, the vortex lattice enjoys full hexagonal symmetry.  Near $H_{c1}$, however, the background superconducting order parameter defines a special nematic orientation, and the well-isolated topological excitations should form a hexagonal array distorted along the nematic direction.  If PVs are favored, the resultant array has $C_{2v}$ symmetry.  If HQV pairs are preferred, the dipolar nature of the pair further reduces the symmetry down to only one mirror reflection.  See Fig \ref{lowField}.  We conclude that the hexagonal symmetry near $H_{c2}$ must be broken in a phase transition as the field is lowered.

Let us denote the high-field hexagonal phase as the $T$ phase, and the two low-field phases as $PV$ and $HQV$ phases, respectively.  Despite having the correct total topological charge, the $\eta_+$ component in the $T$ phase has too many zeroes (three) per unit cell, two positive-winding and one negative-winding.  The $PV$ and $HQV$ phases each has only one negative-winding zero per unit cell.  As the applied field is lowered, pair-annihilation must occur.  We argue that at least two phase transitions must take place during the process.

The locations of zeroes of the $\eta_+$ component are protected by the hexagonal symmetry in the $T$ phase: the negative-winding ones sits at the lattice points, and the positive-winding ones sit at the center of the regular triangles.  A symmetry-breaking transition is required to allow these zeroes to drift.  Eventually the drifting zeros will meet and pair-annihilate, and this must be another phase transition.

\begin{figure*}
\subfloat[][]{\includegraphics[width=0.3\textwidth]{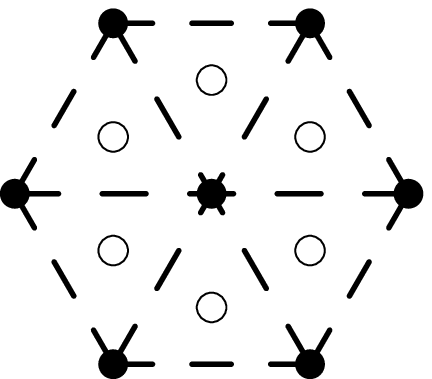}}
\qquad
\subfloat[][]{\includegraphics[width=0.25\textwidth]{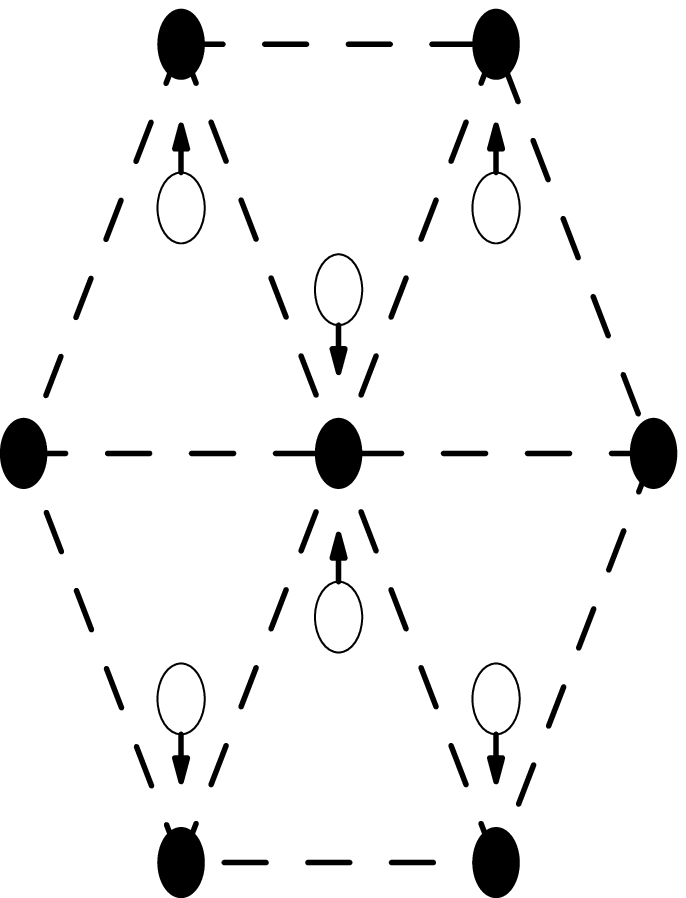}}
\qquad
\subfloat[][]{\includegraphics[width=0.25\textwidth]{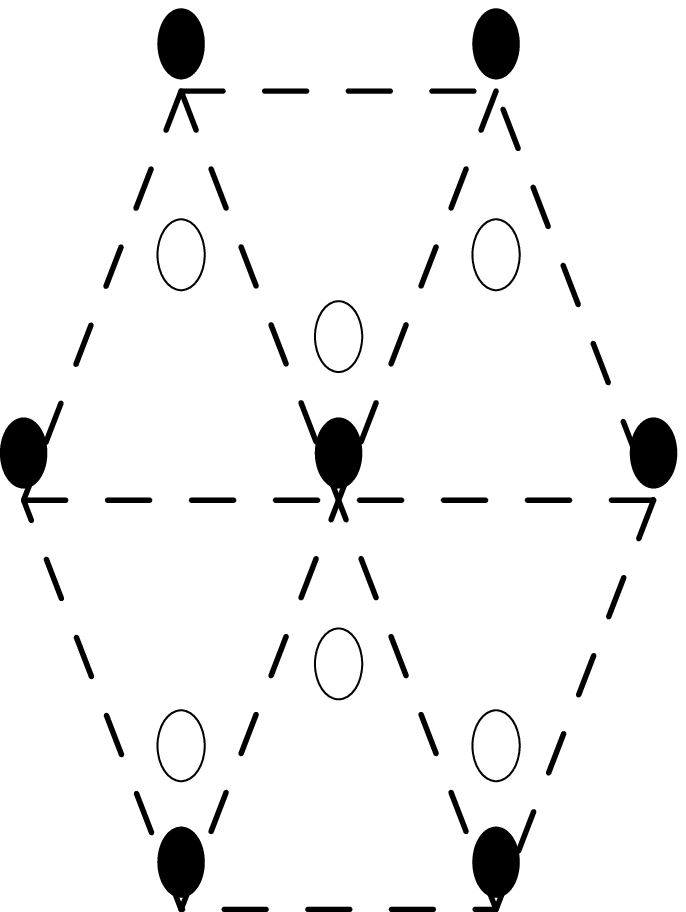}}
\caption{\label{higherField} Illustrations of the phases $T$, $D_1$ and $D_2$.  The vertices of the dashed triangles mark the locations of zeros of the $\eta_-$ component in each figures.  (a) The hexagonal phase $T$.  The circles marks the zeros of the $\eta_{+}$ component; the filled ones are positive-winding, while the unfilled ones are negative-winding.  (b) The $D_1$ phase.  The unfilled circles drift away, breaking the hexagonal symmetry, and the distortion of the background hexagon is also indicated.  The unbroken symmetry group is $C_{2v}$.  (c) The $D_2$ phase.  The filled circles also move, and the only remaining symmetry is the $x \rightarrow -x$ reflection.}
\end{figure*}

We construct a plausible minimal scenario with two intermediate phases $D_1$ and $D_2$.  In either phase, the hexagonal symmetry is distorted along one direction.  $D_1$ still retains a two-fold rotational symmetry, but $D_2$ has only a mirror plane.  See Fig \ref{higherField}.  For $D_1$, eventually all three $\eta_+$ zeros collapse onto the same location of the $\eta_-$ zero, making a single PV per unit cell.  For $D_2$, though, an HQV is left in each unit cell.  The $D_1$ and $D_2$ phases are naturally connected to the $PV$ and $HQV$ phases, respectively.

\begin{figure}
\includegraphics[width=0.35\textwidth]{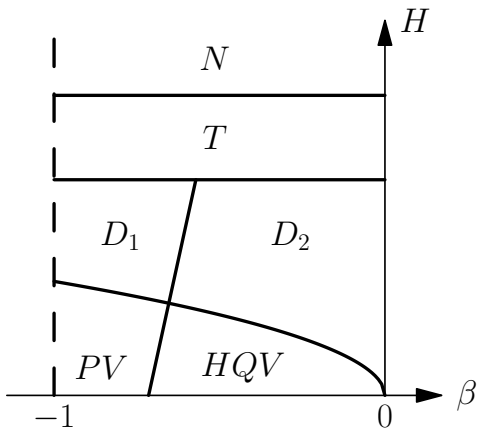}
\caption{\label{beta-H} A plausible phase diagram in the $\beta$-$H$ plane.  $N$ stands for the normal phase, and other phases are defined in the text.  The low end of the $H$ axis schematically represents $H_{c1}$.}
\end{figure}

The phase diagram for our scenario is schematically sketched in Fig \ref{beta-H}.  The phase boundary $T$-$D_1$ and $T$-$D_2$ sees the breaking of three-fold rotational symmetry, among other things.  This is most likely a first order transition\cite{Landau1980}.  There is no reduction in symmetry across $D_1$-$PV$ and $D_2$-$HQV$ phase boundary, so the transition too can only be first order.  The breaking of two-fold rotation across the $D_1$-$D_2$ and $PV$-$HQV$ boundary means the transition may be second order.  Finally, even at low field the $HQV$ phase should cease to exist when $\beta$ changes sign, because it only makes sense in a nematic background.

When $\beta$ is positive, the uniform chiral superconducting state breaks the time-reversal symmetry.  Near $H_{c1}$, it has been shown that the isolated ``crescent vortex'' solution is dipole-like and has only a reflection symmetry\cite{Tokuyasu1990}.  In our proposed $D_2$ phase, the positive-winding zero of $\eta_+$ and the zero of $\eta_-$ can pair up to form a dipole-like structure, while the remaining zeros of $\eta_+$ can broaden out so that the interior of the unit cell between lattice points sees a background strongly dominated by the chiral $\eta_-$.  In other words, our $D_2$ phase can evolve smoothly into a crescent vortex lattice in a chiral background on the $\beta > 0$ side.

\section{Conclusion}  In this paper we study the topological excitations of a nematic superconductor.  At low applied field, an isolated excitation is either a PV or a tightly-bounded HQV pair.  This is in agreement with existing numerical simulation\cite{Zyuzin2017} that observed tightly-bounded HQV pairs at low field.  Available images of vortices in Cu$_x$Bi$_2$Se$_3$\cite{Tao2018} shows only blurry, elongated defects.  It would be interesting if individual HQVs can be visualized in experiments.  For instance, local density of states probed by STM may show a twin-peak structure due to the HQV pair.

We construct the phase diagram for the vortex lattice state near $H_{c2}$.  For the range of parameters relevant to $M_x$Bi$_2$Se$_3$, the vortex lattice is always hexagonal.  This agrees with the finding in \cite{Lukyanchuk1993}.  The vortex lattice solution at $H_{c2}$ developed in this letter is the full solution for an unconventional superconductor without extra simplifying assumptions.  Our method can be adopted for other unconventional superconductors, and may prompt the reexaminations of prior results.

\begin{acknowledgments}
This work was supported by Academia Sinica through AS-iMATE-109-13, and the Ministry of Science and Technology (MOST), Taiwan through MOST-107-2112-M-001-035-MY3.  Additionally, PTH is supported by MOST through MOST-109-2119-M-007-005.
\end{acknowledgments}

\bibliography{HQV}

\appendix

\section{GL parameters and length scales}

In the s-wave case, the GL parameter $\kappa_{\text{s-wave}}$ plays a double role.  On one hand, it is the ratio of magnetic penetration depth to the coherence length, i.e. $\kappa_{\text{s-wave}} = \lambda/\xi$; on the other, it is the sole parameter that distinguishes between type-I and type-II behavior, and vortices are only allowed if $\kappa_{\text{s-wave}} > 1/\sqrt{2}$.  For the present model, the two roles are no longer played by the same quantity.

The criterion for type-II superconductivity is the existence of an upper critical field $H_{c2}$ higher than the thermodynamic critical field $H_{c}$.  For the present theory, $H_{c2}$ was calculated in Ref [29].  In our notation it reads:
\begin{equation}
H_{c2} = \left(\frac{\bar{\kappa}}{\bar{\lambda}}\right)^2
	\left[3(1+C)-2\sqrt{3C^2+2C+1}\right]^{-1},
\end{equation}
where we have introduce
\begin{equation}
\bar{\kappa}^2 = \frac{\beta_1}{8\pi K_1^2}, \qquad
\bar{\lambda}^2 = \frac{1}{8\pi K_1} \frac{\beta_1}{\vert\alpha\vert}.
\end{equation}
$H_c$ can be easily read off of the free energy (1) in the main text.  It is different for chiral and nematic phases.
\begin{equation}
H_c = \begin{cases}
\bar{\kappa}/(\bar{\lambda}^2 \sqrt{2} )
\qquad & \beta > 0, \text{chiral} \\
\bar{\kappa}/ (\bar{\lambda}^2 \sqrt{2(1+\beta)})
\qquad & \beta < 0, \text{nematic}.
\end{cases}
\end{equation}
The requirement of $H_{c2} > H_c$ is captured by
\begin{equation}
\kappa \equiv \frac{\bar{\kappa} \, \min(1, \sqrt{1+\beta}) }{3(1+C)-2\sqrt{3C^2 + 2C + 1}} > \frac{1}{\sqrt{2}}.
\end{equation}

This $\kappa$ parameter, however, is not directly accessible experimentally.  The ratio of length scales $\lambda/\xi$ is the quantity reported in experiments.  The experimental estimations put this ratio typically at the order of $100$ for $M_x$Bi$_2$Se$_3$.  We adopt the s-wave-inspired
\begin{equation}
H_{c2} = \Phi_0/2\pi\xi^2 = \xi^{-2}
\end{equation}
for estimating $\xi$, as is usually done in the literature.  Penetration depth $\lambda$ is extracted from the GL equation for $\vec{A}$ expanded around the uniform state solution.  It is anisotropic in the nematic phase, since the uniform background of order parameter itself breaks the rotational symmetry.
\begin{equation}
\left(\frac{\lambda}{\bar{\lambda}}\right)^2
= \begin{cases}
1/(1+C), &\qquad \text{chiral} \\
(1+\beta)/(1+2C), &\qquad A_{\parallel}, \text{nematic} \\
(1+\beta), &\qquad A_{\perp}, \text{nematic} \\
\end{cases} 
\end{equation}

\begin{widetext}
We define the ratio of length scales
\begin{equation}
\kappa_{r}^2 \equiv \left(\frac{\lambda}{\xi}\right)^2
=
\begin{cases}
\bar{\kappa}^2 / \left\lbrace
	(1+C)\left[3(1+C)-2\sqrt{3C^2+2C+1}\right] \right\rbrace,
	& \qquad \text{chiral} \\
\bar{\kappa}^2 (1+\beta)/ \left\lbrace
	\sqrt{1+2C}\left[3(1+C)-2\sqrt{3C^2+2C+1}\right] \right\rbrace,
	& \qquad \text{nematic} \\
\end{cases}
\end{equation}
For the nematic phase, the geometric mean of the two anisotropic penetration depth is used in the ratio.
\end{widetext}

Assuming $1+\beta$ is not extremely small, as far as order of magnitude is concerned, one can see that $\kappa_r \approx \kappa$ for the entire range of $-1/3 < C < 1$ in the nematic phase.  The above discussion informs our choice of range of $\kappa$ in the vortex lattice calculation based on the experimentally reported $\kappa_r$.

\section{Integration of Maxwell Equation for $h_s$}

In this section we provide the derivation of the result \eqref{magneticField}.  Essentially one wants to undo the curl in \eqref{Maxwell}, while the left hand side is given using the ansatz \eqref{ansatz}.

For the s-wave case, it was originally argue that $h_s \propto \vert \psi_0 \vert^2$ using special identities available only because the linearized solution is the lowest Landau level.  For the present case, the ansatz \eqref{ansatz} contains both $\psi_0$ and $\psi_2$, but the Landau level structure still allows us to integrate \eqref{Maxwell} analytically.

From \eqref{Maxwell} one may write down
\begin{equation}
\partial_y h_s = 4 \pi \frac{\partial \mathcal{F}_2}{\partial A_x} \equiv 4\pi j_x.
\end{equation}
Integrating with respect to $y$, in principle, undoes the partial derivative on $h_s$.  To determine the constant of integration, we replace the Landau level wavefunctions $\psi_j$ defined in \eqref{LandauLevel} with the regularized $\tilde{\psi}_j$, where the infinite sum over $n$ in the original definition of $\psi_j$ is instead cut off at $\pm N$ for some $N$.  The new $\tilde{\psi}_j$ vanishes exponentially as $y \rightarrow \pm \infty$, but remains a $j$th Landau level wavefunction.  With the regularized ansatz, the system is completely normal at $y \rightarrow \pm\infty$, and $h_s$ must vanish there.  We therefore choose to begin the integration from $y = -\infty$.  After the integration is done, the cutoff $N$ is sent to infinity to restore the spatially extended periodic ansatz.  Let $\tilde{\jmath}_x$ denotes the regularized version of $j_x$, and the above discussion amounts to
\begin{equation}
h_s = \lim_{N\rightarrow \infty} 4\pi \int_{-\infty}^{y}\! dy \, j_x.
\end{equation}

The exact form of the current $j_x = \delta \mathcal{F}_2/\delta A_x$ is:
\begin{widetext}
\begin{equation}
j_x = \frac{K_1}{2}\bpm \eta_+^{*} &  \eta_-^{*} \epm
\bpm
\frac{1}{2}(1 + C)(p_+ + p_-) & Cp_+ \\
Cp_- & \frac{1}{2}(1+C)(p_+ + p_-)
\epm
\bpm \eta_+ \\ \eta_- \epm + (\text{c.c}).
\label{current}
\end{equation}
\end{widetext}
Utilizing the commutator $[p_+, p_-] = 2 H_{c2}$, and the Landau levels' property
\begin{equation}
\begin{split}
\tilde{\phi}_n &= \left(\frac{p_+}{\sqrt{2H}}\right)^{\! n} \tilde{\psi}_n \\
p_- \tilde{\psi}_0 &= (p_x - i p_y) \tilde{\psi}_0 = 0, \\
p_+ \tilde{\psi}_0 &= (p_x + i p_y) \tilde{\psi}_0
	= 2 i p_y \tilde{\psi}_0 = 2 \partial_y \tilde{\psi}_0,
\end{split}
\end{equation}
one may trade all occurrences of the ladder operators $p_\pm$ in \eqref{current} for some polynomial of partial derivative $\partial_y$.  The integration with respect to $y$ is now trivial.  The $y \rightarrow -\infty$ part vanishes due to our regularization.  And then the same trick is employed again to trade all $\partial_y$ in the result for $p_+$.  Finally the cutoff $N$ is allowed to go to infinity again.  At the end one recovers \eqref{magneticField} as required.

\section{Helmholtz Free Energies}

So far we have framed the discussion exclusively in terms of the Gibbs free energy, as we deem it conceptually more appropriate for the present case with an externally applied magnetic field.  Historically, however, the procedure has been carried out in terms of the Helmholtz free energy.  We will apply Legendre transformation and establish the equivalence between our result and the conventional lore.

Let $B = \langle \nabla\!\times\!\vec{A} \rangle$ be the average magnetic flux density, and $\Delta B = B - H_{c2}$ will be the new expansion parameter.  Recall that our Gibbs free energy density was defined as
\begin{equation}
\mathcal{G} = \dots + \frac{1}{8\pi}(\nabla \! \times\! \vec{A} - H)^2,
\label{GibbsDef}
\end{equation}
with the addition of the constant $H^2/8\pi$.  To get the usual Helmholtz free energy $F$, one needs
\begin{equation}
\begin{split}
\langle \mathcal{F} \rangle &= \langle \mathcal{G} \rangle + \frac{1}{8\pi} (2BH - H^2) \\
&= \langle \mathcal{G} \rangle - \frac{1}{8\pi} (\Delta B - \Delta H)^2
	+ \frac{B^2}{8\pi}.
\end{split}
\label{Helmholtz}
\end{equation}

Definition \eqref{GibbsDef} also implies that $\partial \langle \mathcal{G} \rangle / \partial H = -(B-H)/4\pi$.  Using \eqref{Gibbs} for $\langle \mathcal{G} \rangle$, one may write down:
\begin{equation}
(B - H) = (\Delta B - \Delta H)
= \Delta H \left(\frac{\langle h_s \rangle^2/8\pi}{\langle \mathcal{F}_4\rangle - \langle h_s^2\rangle/8\pi} \right).
\label{B&H}
\end{equation}
This implies $\Delta H$ and $\Delta B$ are of the same order.  One can now identify from \eqref{Gibbs} and \eqref{B&H}
\begin{equation}
\langle \mathcal{G} \rangle = -\frac{\Delta H}{8\pi} (\Delta B - \Delta H),
\end{equation}
which turns \eqref{Helmholtz} into
\begin{equation}
\langle \mathcal{F} \rangle = -\frac{1}{8\pi} \Delta B (\Delta B - \Delta H)
+ \frac{B^2}{8\pi}.
\end{equation}
Finally, by using \eqref{B&H} again, $\Delta H$ can be eliminated in favor of $\Delta B$.  The end result is in agreement with the existing literature:
\begin{equation}
\begin{split}
\langle \mathcal{F} \rangle &= -\frac{\Delta B^2}{8\pi}
\left(
\frac{\langle h_s \rangle^2/8\pi}
{\langle \mathcal{F}_4 \rangle
	- \langle h_s^2 \rangle/8\pi + \langle h_s \rangle^2/8\pi}
\right) + \frac{B^2}{8\pi} \\
&= -\frac{1}{8\pi} \frac{\Delta B^2}{R+1} +\frac{B^2}{8\pi}.
\end{split}
\end{equation}
Our $R$ indeed generalizes the same quantity minimized in the Abrikosov method, traditionally derived using Helmholtz free energy $\mathcal{F}$.

Finally we note that, for the s-wave case with a scalar order parameter $\psi$, integrating the current yields simply $h_s \propto \vert \psi\vert^2$, while $\langle\mathcal{F}_n\rangle \propto \vert \psi\vert^n$ for both $n=2,4$.  One instantly recovers the famous s-wave result due to Abrikosov:
\begin{equation}
R \propto \frac{\langle \mathcal{F}_4 \rangle}{\langle \mathcal{F}_2 \rangle^2}
\end{equation}

\section{Phase Diagram Fig \ref{beta-H}}

We present a detailed description of the phase diagram FIG. \ref{beta-H} from the main text.  Generally speaking, at fixed $C \neq 0$ and a fixed $\kappa$ that allows for type-II behavior, increasing the value of $\beta$ takes the system through three distinct regimes.  The hexagonal and the square phases are self-explanatory; the transition between them appears sharp in our numerical work.

The irregular regime perhaps should not be called a ``phase'' in the true sense of the word.  One enters this regime by further increasing $\beta$ from the square phase.  At the onset of this regime, the square lattice is deformed into a rectangular one.  And then the free energy developed multiple local minima competing with the rectangular lattice.  These minima have no obvious symmetry other than the oblique one, and their positions on the $(\rho, \alpha)$ plane drift around when the parameters $C$, $\beta$, and $\kappa$ are changed.  The relative differences in free energy between these local minima are typically smaller than $O(10^{-14})$.  We can't resolve the true global minimum with the available numerical precision, but with such a small relative difference, the true ordering is unlikely to have physical significance.

A small $\vert C \vert$ pushes the transitions to larger $\beta$.  This can be understood since the $C = 0$ axis is always in the hexagonal phase.  (At $C=0$, $\eta_+$ and $\eta_-$ decouple in the linearized GL equations, making the problem formally identical to the single-component case, which always favors the hexagonal lattice.)  A larger $\kappa$ also pushes the transitions to larger $\beta$.

As depicted in Fig 3, $\beta_s$ and $\beta_i$ denote the onset of square phase and irregular regime respectively at $C = 1$.  At $\kappa \rightarrow 1/\sqrt{2}$, we find $\beta_s \approx -0.20$ and $\beta_i \approx 0.05$: a nematic superconductor ($\beta < 0$) with low enough $\kappa$ may exhibit square vortex lattice near $H_{c2}$.  However, for $\kappa = 100$, we find $\beta_s \approx 0.32$ and $\beta_i \approx 0.36$, setting $M_x$Bi$_2$Se$_3$ firmly in the hexagonal phase.

We speculate that the irregular regime found here may be related to the reported complicated and disordered vortex arrangement in Sr$_2$RuO$_4$[52], which is described by a similar GL theory with $\beta > 0$.  Its $\kappa$ is conventionally estimated to be a moderate $2.6$[53], favoring the irregular regime according to our finding.  However we note that, precisely in the context of high field vortex arrangements of Sr$_2$RuO$_4$, Ref [52] disputes the validity of the $\kappa = 2.6$ description.

\section{Vortex lattice near $H_{c1}$}

In this appendix, we explicitly argue that the topological excitations form a distorted hexagonal array near $H_{c1}$.  This is regardless of whether these excitations are single PVs or HQV pairs.

Let us first revisit the ansatz \eqref{HQVform} and \eqref{pairForm} for isolated PV and HQV pair, respectively.  It is clear that in either case the orientation angle $\theta$ is roughly a constant at distance $R$ much greater than core size $\xi$:
\begin{equation}
\theta = 0 + O(g) + O(\xi/R),
\end{equation}
where the $\xi/R$ term is the dipolar distortion that simply vanishes for PV.  We have taken the liberty to re-align the coordinate axes so that $\theta_{\infty} = 0$.  Consequently, the unit vector $\hat{u}_0 \approx \hat{x}$, and $\nabla \theta \approx 0$ far away from the topological defect.

Now we re-derive the effective free energy in the extreme London limit, with the gauge sector included, using the above approximation.  The free energy reads:
\begin{equation}
\begin{split}
\mathcal{F} &\sim
\vert\nabla\chi + \vec{A}\vert^2 + g \left[\hat{x}\cdot(\nabla\chi +\vec{A})\right]^2 \\
&\qquad - g \left[\hat{y}\cdot(\nabla\chi+\vec{A})\right]^2
+\frac{M}{8\pi}(\nabla\!\times\!\vec{A})^2,
\end{split}
\end{equation}
where $M$ is the appropriate constant coefficient.  By rescaling
\begin{equation}
\begin{split}
\tilde{x} = x/\sqrt{1+g} , &\qquad \tilde{y} = y/\sqrt{1-g} \\
\tilde{A}_x = A_x\sqrt{1+g} , &\qquad \tilde{A}_y = A_y \sqrt{1-g}
\end{split}
\label{anisotropicScaling}
\end{equation}
one brings the free energy into an effectively isotropic form
\begin{equation}
\tilde{\mathcal{F}} \sim
\vert\tilde{\nabla} \chi + \tilde{A} \vert^2
+ \frac{M}{8\pi} (\nabla\!\times\!\tilde{A})^2.
\label{isotropic}
\end{equation}

It is a well known fact that, for this isotropic theory, single PVs at finite density arrange themselves into a regular hexagonal array.  Assuming $g > 0$ (and consequently $C>0$), to account for the anisotropic scaling \eqref{anisotropicScaling}, the hexagonal lattice and the vortices themselves are elongated in the direction parallel to the background $\vec{\eta}$, and compressed in the orthogonal direction.  When $g < 0$, the distortion is the exact opposite, and this is consistent with the duality where $g$ is mapped to $-g$ if $\vec{\eta}$ is rotated by 90 degrees.  We expect hexagonal vortex lattice and the background $\vec{\eta}$ to both be aligned to some high-symmetry direction of the underlying trigonal lattice of $M_x$Bi$_2$Se$_3$.

We shall assume $g > 0$ in all the subsequent discussion.  The solution \eqref{HQVform} indicates that the shape of an isolated PV receives a quadrupolar distortion due to the background.  More generally, the background (even allowing for periodic modulation due to the vortex lattice) is always invariant under $\pi$ spatial rotation followed by $\pi$ phase shift.  Therefore a distorted PV retains the symmetry of an ellipse, with its semi-major axis aligned with the background nematic direction.  The symmetry point group of this low-field PV state is $C_{2v}$, containing a two-fold axis going through the ellipse's center, and two mirror planes perpendicular to the ellipse's axes.  We will denote this as the $PV$ phase.  See Fig \ref{lowField} for an illustration.

In the discussion of lattice symmetry, any mirror reflection is implicitly followed by time-reversal to compensate for the reversal of winding.  Furthermore, the two-fold rotation and one of the reflection are defined with a global phase shift of $\pi$ to compensate for the reversal of the nematic orientation angle $\theta$.  

Suppose the HQV pair is preferred instead.  The pair is dipolar in nature, but at a distance $R$ from its center, the dipole behavior is suppressed by the small ratio $\xi/R$.  At low density, these pairs should likewise arrange themselves into a distorted hexagonal array.  But the dipole nonetheless breaks the two-fold rotation and the mirror plane not parallel to it.  At the end, there is only one mirror plane left in the symmetry point group.  This phase is denoted as $HQV$.

\end{document}